\documentclass[reprint,pra,leqn,showkeys,showpacs]{revtex4-2}
\usepackage{amsmath,amsfonts,amssymb,mathrsfs,graphicx}
\usepackage[colorlinks,linkcolor=blue,citecolor=blue,urlcolor=blue]{hyperref}
\usepackage[usenames,dvipsnames]{color}

\newcommand\p{\pi}

\newcommand\f{\phi}

\renewcommand\j{\psi}

\def\ket#1{\lvert#1\rangle}

\def\bra#1{\langle #1 \rvert}

\def\amp#1#2{\langle #1 \lvert #2 \rangle}

\def\avg#1#2#3{\langle #1 \lvert #2 \lvert #3 \rangle}

\begin{document}
	
\title{Response to the comment on ``Do Bloch waves interfere with one another ?"}
	
\author{Vivek M. Vyas}
\affiliation{Indian Institute of Information Technology Vadodara, Government Engineering College, Sector 28, Gandhinagar 382028, India\\Email: vivek.vyas@iiitvadodara.ac.in}

\begin{abstract}
Here a generalised argument showing the existence of the Bloch superselection rule is presented, in response to the recent comment by Sowinski. In light of the role played by the periodic boundary condition, locality and topology in the system, the claim made in the comment is found untenable.  
\end{abstract}
%\date{\today}
%\keywords{Topology, Bloch states, Superselection}
\maketitle

It is a well known fact that in quantum physics, physical observables are required to be self adjoint linear operators in the Hilbert space of the given system \cite{ecgbook}. Since there are infinitely many self adjoint operators defined over a given Hilbert space, one wonders if each of them correspond to a physical observable. A careful thought will convince the reader that in general all self adjoint operators will not be physical observables. For example self adjoint operators $m \mathbf{1}$ (where $m$ is real and $\mathbf{1}$ is identity operator) are clearly not physical observables. In the last century after a lot of debate, it was eventually found that in certain quantum systems, if a given self adjoint operator obeys certain algebraic properties then only it can be a valid physical observable \cite{giulini2016superselection,wightman1995superselection}. In such a case one says that the system admits a superselection rule, which partitions the Hilbert space of the system into superselection sectors. As a result there exists no valid observable which connects any two states belonging to different superselection sectors.

In an earlier work \cite{vmv2021}, it was found that the quantum periodic lattice system admits the existence of a superselection rule, which was christened as the Bloch superselection rule. It was shown that this superselection essentially forbids the interference between any two Bloch states $\ket{\j_{n k_j}}$ and  $\ket{\j_{m k_l}}$ with different wave vectors $k_{j} \neq k_{l}$ \footnote{Throughout this note we have followed the notations and conventions of Ref. [1].}. In the treatment it was assumed that a quantum observable $\mathcal{A}$ must be a self adjoint local linear operator which obeys cell periodicity $\mathcal{A}(x,p) = \mathcal{A}(x+a,p)$.

A recent comment on this work by Sowinski \cite{sow}, refutes the existence of Bloch superselection rule by proposing a possible observable $\hat{\mathcal{O}}_{n} = \ket{W^{n}_{M}} \bra{W^{n}_{M}}$ using the
Wannier states $\ket{W^{n}_{M}}$ defined as: 
\begin{align} 
	\ket{W^{n}_{M}} = \frac{1}{\sqrt{N}} \sum_{l} e^{- i k_{l} M a} \ket{\j_{n k_l}},
\end{align}
where $M$ is an integer. It was shown that this proposed observable does not respect cell periodicity, at the same time has nonvanishing matrix element $\avg{\j_{n k_i}}{\hat{\mathcal{O}}_{n}}{\j_{n k_{i'}}}$ for $i \neq i'$, paving the way for interference measurement of the Bloch states with different wavenumbers. So it is claimed in the comment that the existence of Bloch superselection rule is negated in the system at hand.

In order to strengthen and clarify the working of the Bloch superselection rule, as also to negate the claim made in the comment, below we present an alternative derivation of the same based on the edifice of topology, while relaxing the condition of cell periodicity of the observables. It must be mentioned that an intuitive topological argument was indeed presented in Ref. \cite{vmv2021}, and which has not been taken into cognisance in the comment. 

Before we begin the general treatment of Bloch superselection rule, it is pertinent to establish a necessary condition for a self adjoint operator to be considered as a valid observable. We begin by assuming that one is given a periodic lattice system obeying periodic boundary condition and governed by the Hamiltonian $\hat{H} = \frac{\hat{p}^2}{2 \mu} + V(\hat{x})$. In such a system we assume that there exists an observable $\hat{\mathscr{R}}$ such that its matrix elements with the position eigenstates $\ket{y}$ and $\ket{z}$ is some general function $r(y,z)$: $\avg{y}{\hat{\mathscr{R}}}{z} = r(y,z)$. 
Now consider a thought experiment wherein one is able to modify the Hamiltonian of the system to $\hat{H}_{m} = \hat{H} + \hat{\mathscr{R}}$, via some external interaction. In such a case it would be possible to have a transition of the particle completely localised at point $y$ to a distant point $z$ in arbitrarily small time $\varepsilon$ with the finite transition amplitude $\avg{y}{e^{- \frac{i}{\hbar} \varepsilon \hat{H}_{m}}}{z} \simeq \delta(y-z) - \frac{i \varepsilon}{\hbar} \avg{y}{\hat{H}_{m}}{z} = \delta(y-z) - \frac{i \varepsilon}{\hbar} \left(-\frac{\hbar^2 \partial^2}{2 \mu \partial z^2} + V(z) \right) \delta(y-z) - \frac{i \varepsilon}{\hbar} r(y,z)$. Evidently this matrix element has non-trivial contribution only from $r(y,z)$ since we are considering the scenario wherein $|y - z|$ is non-vanishing. Thus one sees that locality is not respected in the system at hand due to the contribution $r(y,z)$, which is problematic, to put it mildly. In case when the particles at hand carry electric charge, this non-locality also implies violation of local charge conservation. 

We know for a fact that neither non-local particle transport nor violation of local charge conservation is experimentally observed in nature, even at subatomic scales. Thus one is only left with the inference that the operator $\hat{\mathscr{R}}$ can not be a valid observable. In other words, a valid observable $\hat{\mathscr{R}}$ must be the one for which the $\avg{y}{\hat{\mathscr{R}}}{z}$ is a local function such that: $r(y,z) = f(z,\partial_{z}) \delta(y-z)$, where $f(z,\partial_{z})$ is some general function of $z$ and $\partial_{z}$. It must be noted that all the well known observables like current, momentum, kinetic energy and potential energy fall in this category, as expected \cite{ecgbook}. It is straightforward to see that such a valid observable respecting locality can always be expressed as a function of $\hat{x}$ and $\hat{p}$. This establishes that \emph{a valid observable must be a self adjoint operator respecting the periodic boundary condition, and be expressible as a function of dynamical operators $\hat{x}$ and $\hat{p}$}. This condition was also spelled in the Footnote in Ref. \cite{vmv2021}. 

In contrast the proposed observable $\hat{\mathcal{O}}_{n}$ is such that the matrix element $\avg{y}{\hat{\mathcal{O}}_{n}}{z} = \frac{1}{N} \sum_{j,l} e^{- i R_{0} (k_{j} - k_{l})} \amp{y}{\j_{n k_j}} \amp{\j_{n k_l}}{z} $, which owing to the extended nature of the Bloch states, is evidently non-local. Thus one learns that the proposed observable $\hat{\mathcal{O}}_{n}$ is invalid for it violates locality. It must be clarified that the invalidity of $\hat{\mathcal{O}}_{n}$ as an observable has no bearing    
on the existence, importance and usage of Wannier states $\ket{W^{n}_{M}}$ for describing physics of the problem at hand. Let us reiterate, the Bloch superselection rule does not impose any restriction or condition on the existence or construction of any linear combination of Bloch states. 

Let us now discuss the general treatment of Bloch superselection rule. 
Consider the action of a local observable $\hat{\mathscr{R}}$ on the state $\ket{\f}$ to yield $\ket{\phi}$: $\ket{\phi} = \hat{\mathscr{R}} \ket{\chi}$. In the position representation this can be expressed as: 
\begin{align} \label{positionrep}
	\phi(y) = \int_{0}^{L} dz \: R(y,\partial_{y}) \delta(z-y) \chi(z),
\end{align}
where $R(y,\partial_{y})$ is some function of $y$ and $\partial_{y}$. While the integration over the cyclic position variable $z$ is from $0$ to $L$, owing to the delta function $\delta(z-y)$, one sees that the non-trivial contribution only comes from the infinitesimal neighbourhood around the point $z=y$. Geometrically the wavefunctions $\phi(y)$ and $\chi(z)$ both respectively represent  closed curves in the complex plane, which are traced when the cyclic position variables $y$ and $z$ are varied from $0$ to $L$. The presence of delta function in (\ref{positionrep}) indicates that in the position representation the action of operator $\hat{\mathscr{R}}$ acts in a local manner so that a point on $\chi(z)$ is mapped to a point on $\phi(y)$, since $	\phi(y) = R(y,\partial_{y}) \chi(y)$. 

Let us now consider the action of such a map $R(y,\partial_{y})$ on Bloch wavefunction $\j_{m k_l}(y)$. As explained in Ref. \cite{vmv2021}, one understands that the Bloch wavefunction $\j_{m k_l}(y)$ itself represents a topologically non-trivial closed curve in the complex plane with the winding number $l$, where $k_{l} = \frac{2 \p}{L}l$. This becomes evident from the fact that $\j_{m k_l}(y) = u_{m k_{l}}(y) e^{i k_{l} y}$, where $u_{m k_{l}}$ is cell-periodic function. 
%From here one sees that the presence of complex exponential function $e^{i k_{l} y}$ is responsible for the non-zero winding number of the Bloch state. 
Given the map $R$ a question arises whether Bloch wavefunction $\j_{m k_l}(y)$ yields some other Bloch wavefunction Bloch wavefunction $\j_{n k_j}(y)$ under the action of this map. In other words, can the action of $R$ connect two closed complex curves with different winding numbers ? A careful reflection reveals that the answer to this question is negative, owing to the fact that the action of map $R$ is local in nature (since $\phi(y) = R(y,\partial_{y}) \chi(y)$). This ensures that an open interval on the former curve is mapped onto an open interval on the latter curve. A well known result in topology forbids the existence of a local map which connects two different topological spaces with different topological index like winding number \cite{nakahara2003}. As a result one infers that the map $R$ due to the observable $\hat{\mathscr{R}}$ can not connect two Bloch states with different winding numbers. In other words, this shows that the non-existence of an observable $\hat{\mathscr{R}}$ which has a non-zero matrix element $\avg{\j_{n k_i}}{\hat{\mathscr{R}}}{\j_{n k_{i'}}}$ for $i \neq i'$.

In fact one can see this result explicitly by realising that in general any local observable $\hat{\mathscr{R}}$ in the position representation can be expressed as a series:
\begin{align*}
	R(x,\partial_{x}) = \sum_{m,n \in \mathbb{Z}^{+}}& \left( c_{mn} \cos ( \frac{2 \pi m }{L}x) (- i \partial_{x})^{n} \right.
	\\ & \left. + d_{mn} \sin (\frac{2 \pi m }{L}x) (- i \partial_{x})^{n} \right). 
\end{align*}
Here the coefficients $c_{mn}$, $d_{mn}$ are real numbers, whereas the indices $m,n$ are positive integers. One can straight away see that the action of $R$ on the Bloch state reads: $\j_{m k_l}(x)$: $R \j_{m k_l}(x) = R ( u_{m k_{l}}(x) e^{i k_{l} x}) = e^{i k_{l} x} F(x)$, where $F(x) (= F(x+L))$ is some periodic function of $x$. The presence $e^{i k_{l} x}$ in the wavefunction $e^{i k_{l} x} F(x)$ shows that its winding number is identical to the Bloch state $\j_{m k_l}(x)$.

Thus we have shown the non-existence of a valid observable $\hat{\mathscr{R}}$ which has a non-zero matrix element $\avg{\j_{m k_i}}{\hat{\mathscr{R}}}{\j_{n k_{i'}}}$ for $i \neq i'$. We have also shown that for $\hat{\mathscr{R}}$ to be a valid observable it must be a local self adjoint operator respecting periodic boundary condition. The proposed observable $\hat{\mathcal{O}}_{n}$ is found invalid owing to its non-local nature. It must be emphasised that the above argument furthering the existence of Bloch superselection rule is a generalisation of the one presented in Ref. \cite{vmv2021}, and is solely based on the grounds of topology.

%\bibliographystyle{unsrt}
%\bibliography{super}	


\end{document}